# Texture Object Segmentation Based on Affine Invariant Texture Detection

Jianwei Zhang[a,*] and Xu Chen[b], Xuezhong Xiao[b]

[a]*Beijing University of Posts and Telecommunications, Beijing, 100876, China*
[b]*Nanjing University of Posts and Telecommunications, Nanjing, 210028, China*

## ABSTRACT

To solve the issue of segmenting rich texture images, a novel detection methods based on the affine invariable principle is proposed. Considering the similarity between the texture areas, we first take the affine transform to get numerous shapes, and then utilize the KLT algorithm to verify the similarity. The transforms include rotation, proportional transformation and perspective deformation to cope with a variety of situations. Then we propose an improved LBP method combining canny edge detection to handle the boundary in the segmentation process. Moreover, human-computer interaction of this method which helps splitting the matched texture area from the original images is user-friendly.

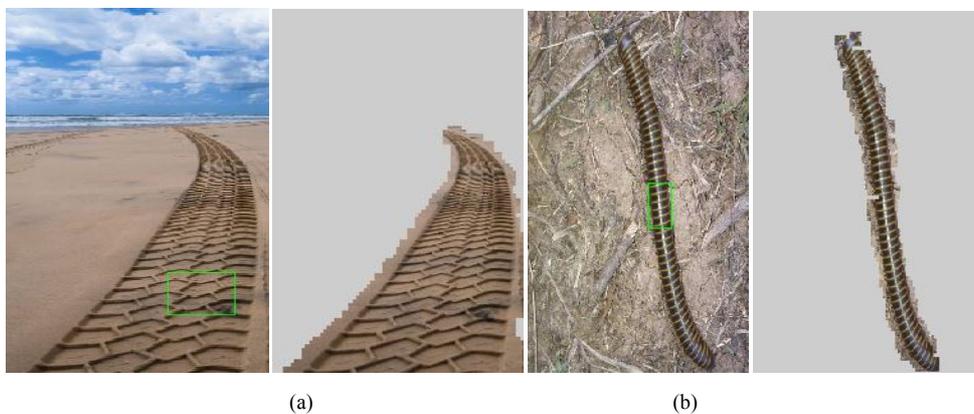

(a)             (b)

Figure 1 Selecting the tiny area and deriving the global similar texture content.

## 1. Introduction

For most applications in the area of image processing and computer vision, image segmentation is an indispensable part. It usually appears in the preprocessing step, such as image recognition and comprehension. And the performance of subsequent steps will be affected by its output enormously. Thus, image segmentation technology has attracted a lot of attention. In the last several decades, there are numerous remarkable achievements, which could be roughly separated into two categories including the image segmentation techniques in the spatial [13,14] and frequency domains [10,20,21]. However, due to the complexity of their backgrounds, none of these technologies, particularly the segmentation techniques aiming at image objects with rich textures, are practical.

In this work, we focus on the segmentation of images with the rich texture regions. To achieve a fast method, we first select a small area as the prior knowledge. Considering the specific region possesses the self-similarity among the texture elements, then we propose a method which can detect the similar texture area based on affine invariant principle. We use the method of the translation, rotation, scale and perspective transformation to detect the texture target area. Meantime, for those images with strong three-dimensional information, the perspective transformation could handle the images well comparing with the affine transformation.

We have got a specific texture region after all the transformations. To make the edge smoother, we first utilize the canny edge detection to get the peripheral line of the texture area. Then enhance the local binary pattern method to make the boundary more accurate. Segmenting automatically is the mainly advantage of our algorithms. We don't need redundant user interactions while processing, while the grab cut algorithms [16] may need more interactions. Extensive experiments on different images including clothing, construction, Knitwear and other rich texture objects show that the proposed algorithms significantly outperforms the state-of-the-art. As we can see, with the trend of online shopping for goods like clothes, the algorithms could be useful to change the texture of the clothes. And the experiment result for the long-time consumption of segmentation is even better.

## 2. Related Work

For those color texture images, Kan-Min Chen [4] has proposed the method of utilizing the feature distribution. It uses color and local edge pattern as the extract texture descriptors to measure the homogeneity for texture region. By quantizing the color image, they then fed it into a color texture segmentation algorithm to segment the image into diff erently textured regions. Qing Zhang

---

* Corresponding author. Tel.: +86-18146539952; e-mail: zhangjianweibupt@bupt.edu.cn

[5] also used the color and texture cues to compute saliency by integrating the background-based and foreground-based patches to present a new bottom-up salient object detection method. Q Zhao [23] has employed the EM variants into a color-texture image segmentation algorithm. like X. Liu [17] and J. Yuan [18] proposed, Local Spectral Histogram is also useful in the Gabor filter. The Gabor filter in the bank is also tuned to detect patterns of a specific frequency and orientation [24]. J Shi [19] has utilized the oriented Gaussian filter in the segmentation. Combining the filters like Cimpoi M has raised the proposal [22] could apply to the texture recognition, description, and segmentation. There may be several distinct areas in a picture, and the clustering algorithms could be a way to solve the segment problem [11, 12, 25].

Though the grab cut algorithms [16] could produce a better segment outcome, the multiple interactions makes operations cumbersome. The automatic segmentation is practicable in some fields like the medical and so on [28-30]. Meanwhile some scholars have put forward applying the machine learning algorithm and deep learning methods into the segmentation [31-37]. We could train the database to segment the interest region by the learning algorithms. But the algorithm is an input-output process which has not considered the content of regions. V. B. Surya Prasath [6] proposed a flexible segmentation approach using multi-channel texture and intensity in a globally convex continuous optimization framework. And the local feature density function is applying to the luminance-chromaticity decomposition.

As we can see, the rich texture area has a high degree of local similarity. The features are invariant to image scaling, translation, and rotation, and partially invariant to affine or perspective transformation. David G. Lowe [2] has previously proposed the basic algorithms SIFT which used the local image features. Wang [3] has redefined the transform map to encode local affine deformations. They chose each block a 2×3 matrix D that maps any image point p=(x, y, 1)$^T$ to position Dp in the epitome. They match the similar patches which were computed previously through affine transformation with the chose block. To measure the degree of self-similarities, they defined and calculated the error e based on the sum of squared differences.

## 3. Algorithm

In this paper, we introduced a new method to search similar contents based on the affine invariant principle. We select a block 'M' as the prior knowledge to derive the similar regions from the original image. To improve the efficiency and practicality of the algorithm, we split up the block into some small blocks named 'mi' (s×s pixels). These small blocks will be as references for the match searching. to find the similar areas (Figure 2), we search each block 'Pij' (s×s pixels) which traversed in the input image (original I) spaced every 3 pixels. In the beginning of the all work, we utilize the color histogram and feature corner point to delete the apparent wrong blocks to exceed the operate rate.

Then, we mainly divide it into two parts. The first step is to detect the similar texture region within the affine invariable principle. We transform the block 'mi' into numerous shapes. And perform match searching by using the Kanade-Lucas-Tomasi (KLT) feature tracker [8], which is widely used to track the which is widely used to track the movement between sequences of frames [Lucas and Kanade 1981; Shi and Tomasi 1994]. Due to the KLT algorithm is based on the basis of small movements, we should initialize the KLT feature tracker with a good state first.

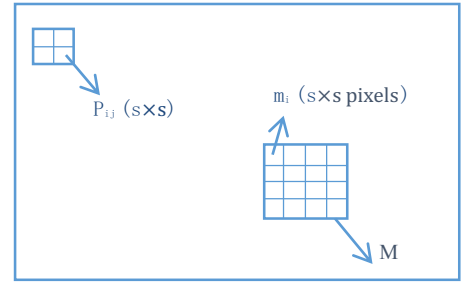

Figure 2. Explanation of some blocks

To express the self-similarity between the two blocks, we define the f(I$_i$, I$_j$) as the value of self-similarity. Mathematically, it can be expressed as Eq. 1:

$$f(I_i, I_j) = \sum_{i}^{i+s} \sum_{j}^{j+s} \| m_i(i,j) - P_{ij}(i,j) \|^2 \quad \text{Eq. 1}$$

Finally, we can segment the texture region based on the detection work. In the detection work we introduced an improved local edge pattern method. And the smooth edge is achieved by combing the canny operator, which is showed in detail in section 3.2. All algorithm flow chart is performed as Figure 3 follows:

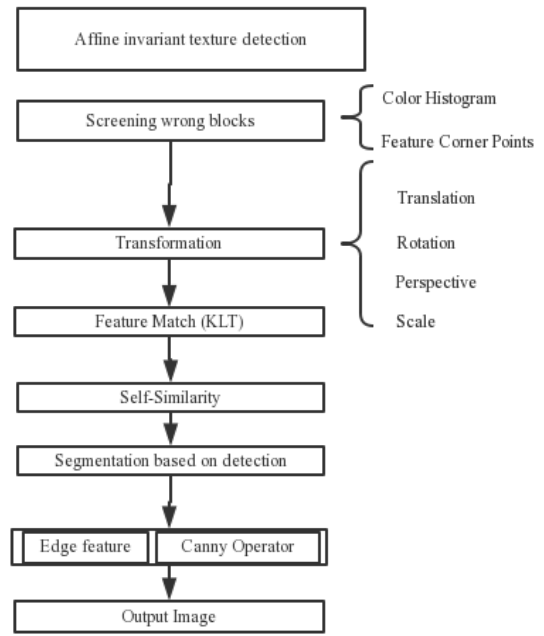

Figure 3. the algorithm flow chart

### 3.1. Texture detection within affine invariable principle

*a. KLT feature match*

To measure whether the two blocks 'mi' and 'Pij' is matched, we use KLT algorithm for match searching. And we detect the feature points in the 'mi' by using the corner detector, initialize separate KLT feature tracker to track these feature points, then judge whether they still exist in frame. We use the corner detector to detect the feature points in the 'mi', and initialize separate KLT feature tracker to track these feature points, then judge whether they still exist in the 'Pij' and record the information of the area around the correspondence points. We stipulate that if the proportion of unmatched feature point is less than 0.15, it is regarded as the matched grid.

*b. Screening based on color histogram and feature corner points*

The first approach is to filter out the areas differentiate from 'mi' apparently, which can be achieved by the HSV color histogram. Compared with the RGB color space, the former can retain more information in the post-processing. So, firstly we compute the HSV color histograms of 'mi' and 'Pij', remove those areas whose adjacent color histograms are too different from 'mi'. Then we have gained the rough area $\emptyset('P_{ij}')$ by traversing the all sets of grids. That means that all 'mi' are corresponds to the area $\emptyset('P_{ij}')$. Meanwhile we skip some blocks according to the divergence in the number of feature points between 'mi' and $\emptyset('P_{ij}')$. And we find that the first screening is necessary and useful for improving the speed. For 'Pij' in the remaining areas, we label it which means it is similar to 'mi'.

*c. Transformation*

We have got the rough region which exists some wrong labels after the first round of screening. To make the results more accurate, we must check every block in $\emptyset('P_{ij}')$ to judge whether it is similar with 'mi'. and we use KLT method to detect the similarity between them. At the same time, many areas in $\emptyset('P_{ij}')$ have been transformed and lost some feature points. We need to make some transformations to 'mi', and then use it as the input of the KLT feature tracker. Not only have we considered some simple transformations, but also complicated transforms like affine transformation and perspective transformation are included. The area obtained by affine transformation and perspective transformation combining proportional transformation as Figure 4 shows.

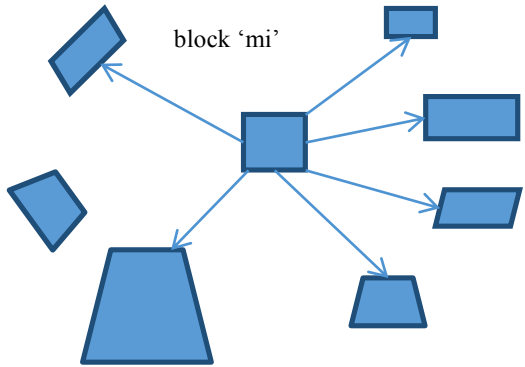

Figure 4. Transformation area

Transform steps as follows:

**Translation and Mirror Reflection** We make the specific transformation for each 'mi' block first, than we use the transformation blocks as the input of KLT to match the remaining area in the original I and mark the similar region (Figure 5). For the mirror reflection, the KLT match result is the same as the translation because it don't change the feature points and fearture points of surrounding area.

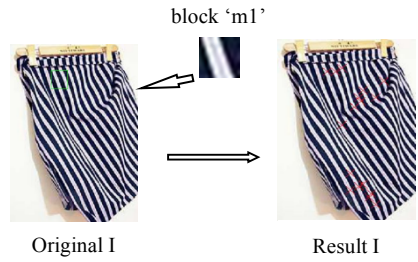

Figure 5. The green rectangles in original I represent block 'M'. The red marks in the Result I' mean that it matches block 'm1'

**Rotation and Scale** We have gained the second round of screening by translation and mirror reflection. But there still be some areas which have not been found. Now we keep on judging 'mi' whether match the leftover grids through rotation and scaling transformation. We not only rotate the 'mi' spaced 5 degrees, but also consider the situation of minification and magnification. As the image pyramid may lose the information at high level, we direct use the scaling within the affine transformation. We only think over the scale ranging between 0.8 and 1.3. And the result is as Figure 6 shows.

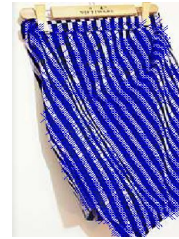

Figure 6. The blue marks mean that it matches block 'mi' which has been transformed by rotation and scale.

**Perspective Transformation** For some images which possess the 3-dimensional information, only considering the affine transformation is not enough. So we introduce the perspective transformation to dispose the problem that affine transformation can't solve. And because the affine transformation is subset of the perspective transformation, we only need to think over the perspective transformation to make it comprehensive. We have considered the main four aspects: bend inward from above, below, left, right direction. Meanwhile, the scaling and rotation is included in every direction. Then we can get the transformed block δ('mi') and use it to initialize the KLT, which is still as a measure to detect δ('mi') whether match the transformed region (Figure 7).

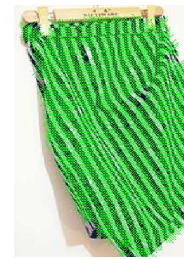

Figure 7 The green mark means that it matches block δ('mi').

In fact, we perform the above mentioned searching operation in turn. While performing in the next round, we jump over the grids labeled in previous processes. If an unlabeled block's neighbors are labeled, it has the similar content in most case and whether it is marked will not play a significant impact on the segmentation. To speed the next round detection, we can label it as right. As the Figure 8 shows, we will label the block in the default way.

## 3.2. Segmentation based on the detection

For those labeled grids which have been verified to possess the similar content with 'mi', we devise a form to record the positions of them and set white to them on a binary image initialized with black, obtaining an obvious dividing line. Then we commence to handle the error areas and the grids near the line.

**Delete Apparent Error Areas**  For some areas which existed detection error like yellow block in Figure 9, we adopt a method in probability to remove them by analyzing the surrounding grid. It is obvious that there are 81 square grids in a result of traversing spaced 3 pixels. We define the rate "α" for the ratio of the grid labeled in all 81 grids to all grids. And once the value of α is less than 0.05. We can assume that the yellow block is a wrong labeled area that we will delete it.

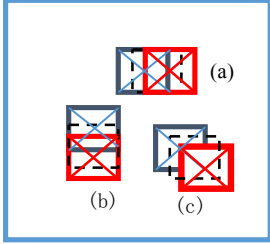 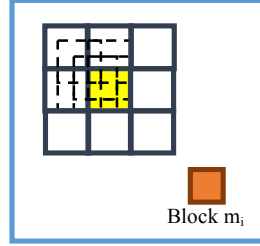

Figure 8 Speeding the detection    Figure 9 Detect error areas

**Edge Feature**  We have got a new idea from the Local binary pattern (LBP) [9] operate on the $3\times 3$ matrixs. The operator calculated the mathematical relationship between the eight neighbors and the central pixel. As we traversal the image by the block 'Pij', it's suitable that Pij could be divided into some $3\times 3$ smaller block. We can assumed that the central pixel as X, the eight neighbors as xi and take an example that the Pij as $12\times 12$ block. And the formula is as Eq. 2:

$$E_{ij} = \sum_{}^{8}(x_i - X)^2 \quad \text{Eq. 2}$$

the value of i and j is 0,1,2,3, where $E_{ij}$ means the values of $3\times 3$ smaller block. It's obvious that the value of $E_{ij}$ will get higher than neighbors [Figure 10.a.b] if the texture edge go through the block. Now it's easy to get the two-dimensional array E [Figure 10.c], we define the $\tau_i$ as the mutation rate which means the amount of change between $E_{ij}$ adjacently in the i-th row.(Eq. 3)

$$\tau_i = \frac{|E_{ij+1} - E_{ij}|}{\max(E_i) - \min(E_i)} \times \frac{|E_{ij+1} - E_{ij}|}{\min(E_{ij+1}, E_{ij})} \quad \text{Eq. 3}$$

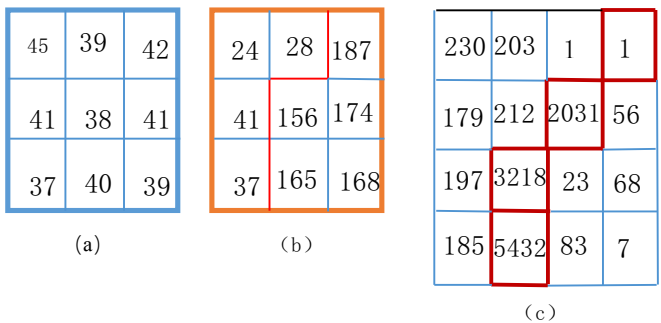

Figure 10 the description of local edge feature

**Edge Processing**  We have got the rough edge of the region, and then we need to make the edge smoothly. We simply use canny edge operator [15] and low-pass filter to achieve it (Figure 11).

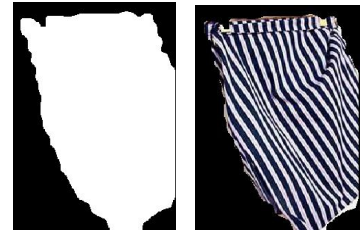

Figure 11 The filter outcome

## 4. Experimental Results

We select the block size of 'mi' as 12×12 pixels, the same for 'Pij'. We have tested our algorithms upon other images with rich texture (see Figure 12). The wall in the first row is from a realistic picture, and is made of countless small bricks. It's easy to draw the whole similar content from the tiny selected region. The textile in the second row is the same way to segment the area of bottom right corner. The third shows two stage zipper which exist some wrong blocks labeled by red crosses in the middle image. It should be that the background is not too complicate in the black edge. But it will disappear a part of wrong blocks in the segment process. We are basically able to separate the coat out at the model who wears a black and white coat picture in the fourth row. The starfish is almost segment well in the fifth row. The cobra in the last row has strong space sense that is possible to detect the texture by perspective transformation.

The detection process is vital to determine the integrity of the texture region, and segmenting the label region is not difficult to work out. Speeding up the process is feasible in the future research, especially for those composite textures images.

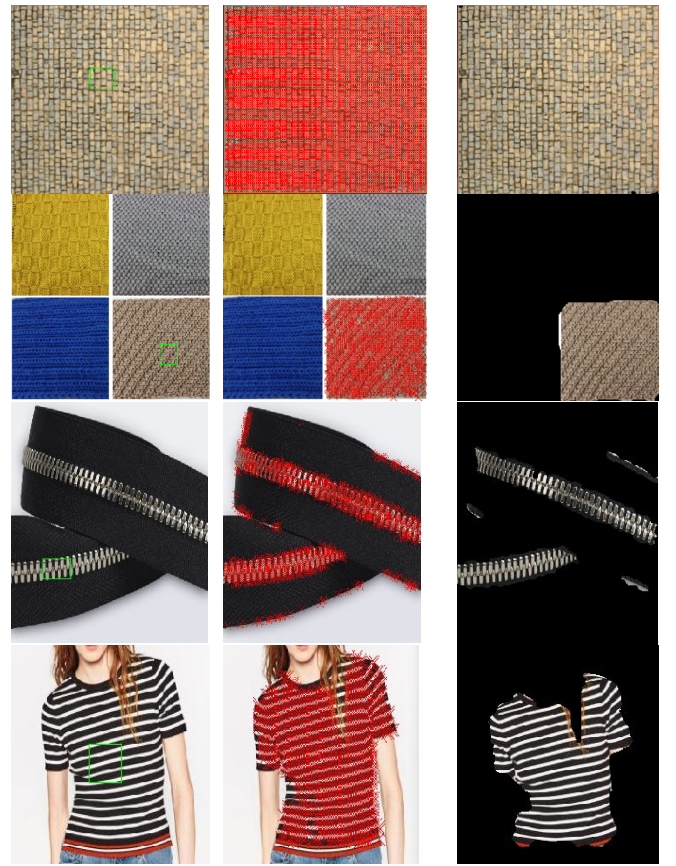

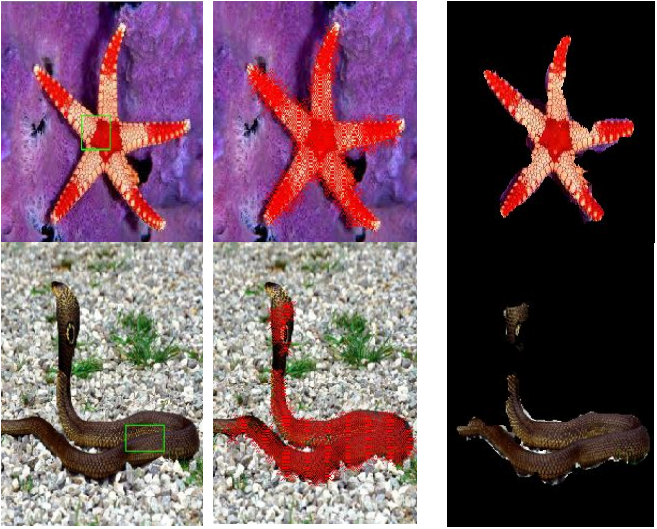

Figure 12 additional segmented images

Meanwhile, we have tested some pictures above by the Grab cut algorithms, the outcome images are performed as follows (Figure 13):

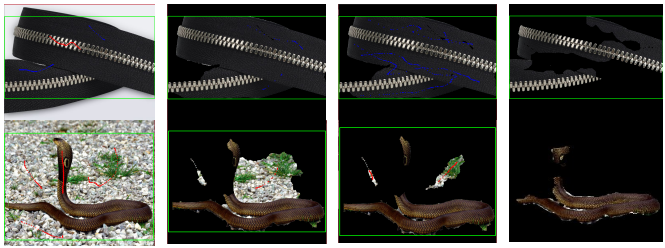

Figure 13 images segmented by grab cut algorithms

Here is the comparison of final outcomes for two algorithms (Figure 14):

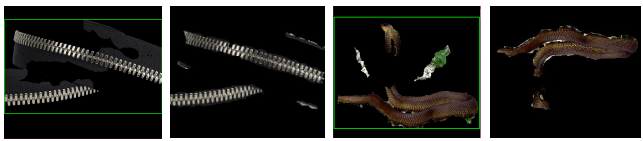

Figure 14 the comparison between the two algorithms

And it's easily find that the traditional segment interaction algorithms, such as the grab cut algorithm, may not get a good effective graph after only one interaction, even the final effect after several interactions is not ideal. Like Figure 13 images segmented by grab cut algorithms, if the zipper and the cobra are completely separated, it will take multiple interactions. As Figure 14 the comparison between the two algorithms shows, the final segment is uncompleted by grab cut. However, our algorithm could achieve better experimental effect as well as higher degree of automation than the grab cut algorithms.

**Block Size**  The graph in Figure 15 explores how the number of blocks selected varies as a function of the block size. Owing to the selection of the size of the block has an effect on the detection. So the comparison of the size of the different blocks makes sense. As the block size increases, the selected blocks get fewer because the content of the block will get more complexity. On the other hand, the number of blocks selected by the perspective transformation is more than other transformation. It's the result of the image containing the 3-dimensional information. We choose the Figure 4(200×300 pixels), set the quantity of the "mi" as 6 as an example and the analysis as follows:

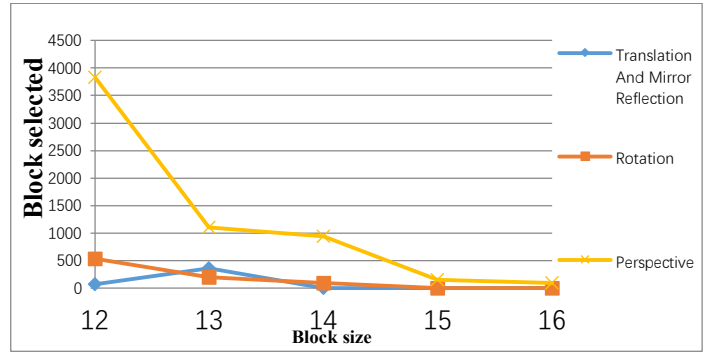

Figure 15 The relationship between the block selected and block size

## 5. Conclusion

We have proposed a novel affine invariant method for detection and segmentation in rich texture images. Our algorithms have following advantages. First, for those rich texture images, we have proposed a new solution which will not be affected by the complex edges and contents compared with previous methods. Second, our algorithms will not be affected by the existence of shadow or light in the background based on the affine invariant. Third, we don't need to do many interactions since it is calculated automatically. However, these algorithms also have limitations, due to traversing and calculating the all sets of grid, the execution time is relative long. The process can be accelerated by parallel-computing with GPU in the future.

## Acknowledgments

This work is supported by the Jiangsu province college student innovation and entrepreneurship plan (program number SYB2016007). And thanks for Shuo Chen`s valuable comments which improves the quality of this manuscript.